\documentclass[aps,prl,reprint,superscriptaddress,nofootinbib]{revtex4-2}

\usepackage{graphicx}      
\usepackage{amsmath}       
\usepackage{amssymb}       
\usepackage{bm}            
\usepackage{dcolumn}       
\usepackage{color}
\usepackage{orcidlink}

\newcommand{\dLt}{\Delta\tilde{\Lambda}}

\begin{document}

\title{How Loud Must a Neutron-Star Merger Be to Reveal Its Equation of State?}

\author{Sk Md Adil Imam\orcidlink{0000-0003-3308-2615}}
\email{adil.imam@unab.cl}
\affiliation{Instituto de Astrof\'isica, Departamento de F\'isica y Astronom\'ia, Universidad Andr\'es Bello, Santiago, Chile}

\date{\today}

\begin{abstract}
The tidal response of neutron stars during binary inspiral encodes the
equation of state (EOS) of dense matter in the gravitational-wave
signal. Quantifying the signal-to-noise ratio (SNR) required to
distinguish competing EOS models with third-generation detectors is
therefore essential. We perform Bayesian nested-sampling parameter
estimation on simulated binary neutron star signals observed by an
Einstein Telescope plus two Cosmic Explorer detector network and
compute the evidence difference between correct- and incorrect-EOS
recovery models over a broad range of SNR. Across two
tidal-deformability contrasts, a swap of the true and recovery EOS,
and two binary mass points, we find a common scaling,
$\Delta\log Z = A\,\mathrm{SNR}^{n}$ with $n \simeq 1.74$--$1.95$,
where the EOS contrast and binary properties determine only the
prefactor $A$. This behavior follows from an Occam-factor argument,
yielding $\Delta\log Z \propto (\Delta\tilde{\Lambda}\,\mathrm{SNR})^{2}$.
Calibrating this relation on three configurations predicts, before the
run, the SNR required for decisive EOS discrimination in the fourth to
within $0.3\%$. These results establish a quantitative framework for
assessing the EOS-discrimination reach of third-generation
gravitational-wave detector networks.
\end{abstract}

\maketitle


\textit{Introduction.}---
The tidal deformation of a neutron star (NS) during binary inspiral
leaves an equation-of-state (EOS) dependent imprint on the
gravitational-wave (GW) phase, encoded to leading order in the
mass-weighted tidal deformability $\tilde\Lambda$~\cite{Read2013}.
GW170817~\cite{Abbott:2017vwq} demonstrated that this imprint is measurable, but
current-generation detectors constrain $\tilde\Lambda$ only weakly;
third-generation (3G) facilities such as the Einstein Telescope (ET)~\cite{Punturo:2010zz}
and Cosmic Explorer (CE)~\cite{Reitze:2019iox} are expected to deliver order-of-magnitude
gains in tidal-parameter precision~\cite{Puecher:2023twf}. Substantial
work has quantified how precisely $\tilde\Lambda$ itself can be
measured as a function of detector sensitivity and astrophysical
population~\cite{Wade2014,LackeyWade2015,Puecher:2023twf}, and recent
work has begun optimizing the Bayesian model-selection machinery
needed to compare discrete candidate EOS models against GW
data~\cite{Kashyap:2025cpd}, alongside related Bayesian
approaches probing degeneracies between the EOS and modified
gravity~\cite{Biswas2023}. What remains largely unquantified—and is more directly relevant for observational planning—is the question addressed in this study: given two candidate EOS models, what SNR must a single BNS merger observation attain to distinguish one EOS from the other?

We answer this with full Bayesian nested-sampling parameter estimation
on simulated BNS signals in an ET + two-CE (ET+2CE) network, computing
the evidence difference $\Delta \log Z$ between correct- and
incorrect-EOS recovery as a function of network SNR. This is a
model-selection statistic, not a parameter-measurement uncertainty:
our results are a statement about discriminating between members of a
finite, named candidate set --- analogous to how LIGO-Virgo-KAGRA's
own GW170817~\cite{LIGOScientific:2019eut} EOS analyses rank a fixed table of nuclear-theory EOS ---
rather than a claim about recovering the true EOS from within a
flexible, continuously parameterized family. We return to this
distinction, and its consequences under model misspecification, in
the Discussion.

Our central result is that $\Delta \log Z$ obeys a near-quadratic
power law in SNR, robust across the tidal-deformability contrast
between the two candidate EOS, which of the two is the true injected
model, and the binary's component masses. This scaling can be understood analytically: a Laplace/Occam-factor expansion of the
evidence integral, together with the established fact that a waveform
mismatch linear in $\tilde\Lambda$ generates an SNR$^2$ evidence
penalty~\cite{Lindblom:2008cm,Read2013}, predicts exactly this
form with a contrast-dependent prefactor. We use this relation to make
a falsifiable, out-of-sample prediction --- calibrated on three
configurations and tested, before the analysis was carried out, on a fourth --- and
confirm it to within a fraction of a percent. We are unaware of previous work demonstrating that the SNR threshold for GW-based EOS
discrimination can be predicted analytically for an unmeasured
configuration, rather than fit only after the fact.

\textit{Method.}---
We perform full Bayesian parameter estimation with the \texttt{jimgw} \cite{Edwards:2023sak,Wong:2023lgb,Wouters:2024oxj}
nested-sampling package, using a BlackJAX-accelerated nested-sampling
annealed-walkers (NS-AW) sampler ($n_{\rm live}=500$), a heterodyned
relative-binning likelihood, and the \texttt{IMRPhenomD\_NRTidalv2}\cite{Dietrich_2017,Dietrich_2019}
waveform. The detector network is ET plus two CE detectors (ET+2CE), representative of a future mature 3G network; sky location is held fixed across all main-text runs. All injections use zero detector
noise, enforcing the exact relation
$\log\mathcal{L}(\theta_{\rm true}) = \tfrac12\,\mathrm{SNR}_{\rm opt}^2$;
we use this both to set the injected distance $d_L$ needed for a
target SNR and as an independent cross-check against the network SNR
obtained by quadrature-summing the per-detector matched-filter SNRs,
which agree to five or more significant figures at every point
reported here.

We construct two NS EOS by hybrid TOV~\cite{Oppenheimer:1939ne, Tolman:1939jz} integration --- a stiffer ``DD2$_{\epsilon_h}$"
model and a softer ``SFHo"
model~\cite{SteinerHempelFischer2013}, differing substantially in
$\tilde\Lambda$ --- plus a third, near-degenerate ``DD2$_{\epsilon_l}$"
variant obtained by tuning the hybrid EOS's transition energy density
to roughly halve this contrast. Both hybrid variants match a speed-of-sound-parameterized high-density grid onto the DD2~\cite{Typel2010} hadronic branch at a transition energy density $\epsilon_t$, with $\epsilon_h=500\,\mathrm{MeV\,fm^{-3}}$ for the fiducial DD2$_{\epsilon_h}$ model and $\epsilon_l=200\,\mathrm{MeV\,fm^{-3}}$ for the near-degenerate DD2$_{\epsilon_l}$ variant.
$\Delta \log Z \equiv \log Z_{\rm correct} - \log Z_{\rm wrong}$ is the
evidence difference between a recovery run whose template EOS matches
the injected EOS and one whose template EOS does not; the two share
identical priors on every parameter except the assumed EOS's
mass-to-$\tilde\Lambda$ mapping.

We report four configurations (Table~\ref{tab:summary}), spanning
three orthogonal variations at fixed network and fixed injected
extrinsic parameters: Curve~1 (DD2$_{\epsilon_h}$ true, SFHo wrong, contrast
$\approx51\%$) is fiducial; Curve~2 (DD2$_{\epsilon_h}$ true, DD2$_{\epsilon_l}$
wrong, contrast $\approx28\%$) isolates reduced tidal contrast at
fixed mass and EOS role; Curve~3 (SFHo true, DD2$_{\epsilon_h}$ wrong) swaps which
EOS is injected and which is recovered, isolating EOS-role symmetry;
and Curve~4 (DD2$_{\epsilon_h}$ true, SFHo wrong, a second independent binary mass
point) tests generalization to a different chirp mass and mass ratio,
and doubles as the out-of-sample test of the analytic prediction
developed below. Each curve spans eight SNR points from
$\mathrm{SNR}\approx15$ to $\mathrm{SNR}\gtrsim3000$, generated by
rescaling only the injected luminosity distance; the highest-SNR point
in each curve uses a narrow Gaussian prior on $d_L$ anchored at the
injected value (all other points use a broad power-law prior) to keep
sampling tractable at extreme SNR. This anchor configuration ($d_L$=42 Mpc) was chosen to coincide with GW170817's inferred distance, with sky position and inclination fixed to their electromagnetic-counterpart-informed values from that event~\cite{LIGOScientific:2017ync,Hotokezaka2019Hubble}; we retain this same fixed sky position and inclination at every other grid point as $d_L$ is increased and the $d_L$ prior reverts to the broad power-law used elsewhere, so that only the loudness (SNR), not the localization quality, changes across each curve.

\begin{figure*}[t]
    \centering
    \includegraphics[width=\textwidth]{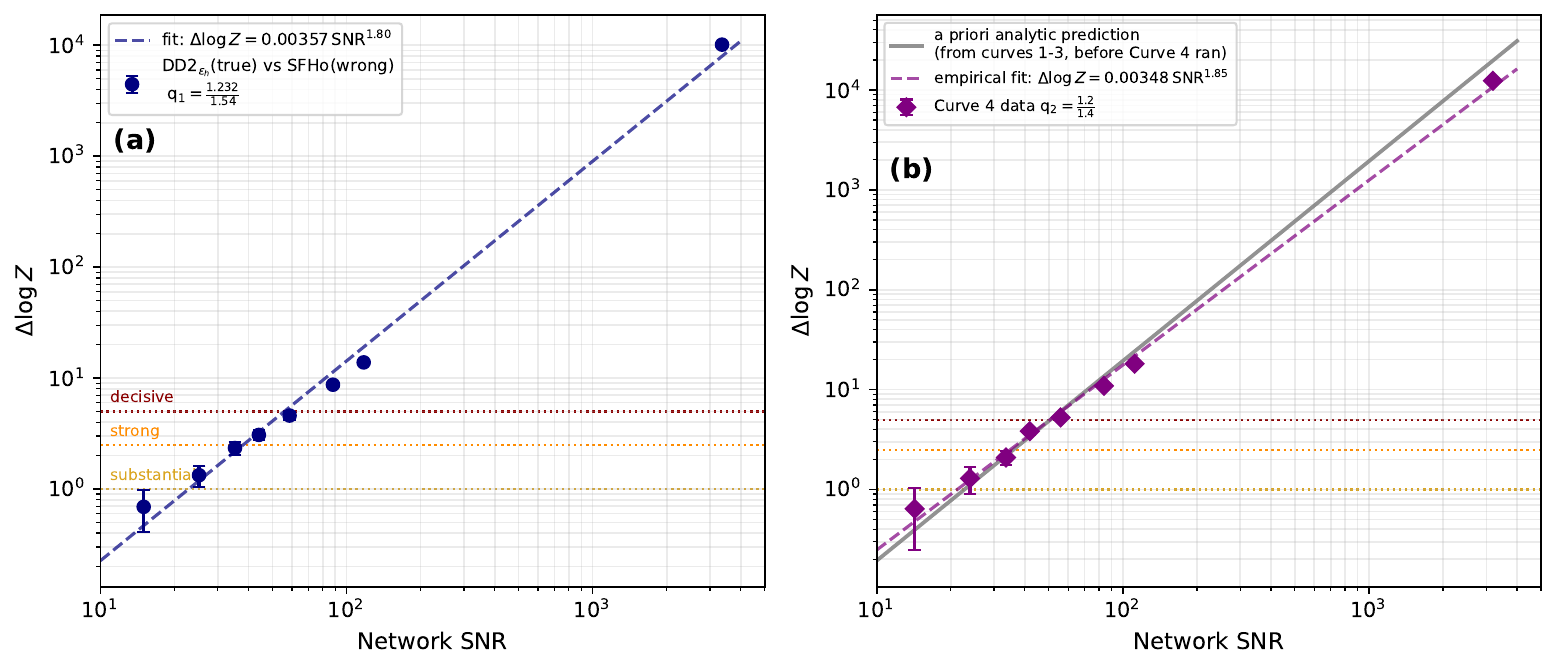}
    \caption{(a) Bayesian evidence difference $\Delta \log Z$ between
    correct- and incorrect-EOS recovery vs.\ network SNR for the
    fiducial configuration (Curve~1: DD2$_{\epsilon_h}$ true, SFHo wrong,
    $\tilde\Lambda$ contrast $\approx51\%$). (b) The same quantity for
    an independent binary mass point (Curve~4), overlaid on the
    \emph{a priori} analytic prediction (gray) obtained from the mean
    Occam-factor coefficient $C$ calibrated on Curves~1--3 \emph{before}
    Curve~4 was run; the empirical fit (purple dashed) and the
    pre-registered prediction agree to within $0.3\%$ at the decisive
    threshold. Dotted horizontal lines mark the Jeffreys'
    substantial/strong/decisive evidence thresholds.}
    \label{fig:delta_logz_snr}
\end{figure*}

\textit{Results.}---Figure~\ref{fig:delta_logz_snr}(a) shows the
fiducial configuration (Curve~1: DD2$_{\epsilon_h}$ injected, SFHo recovered,
$\tilde\Lambda$ contrast $\approx51\%$). $\Delta\log Z$ rises smoothly
with SNR over more than two decades, well described by
$\Delta\log Z = (3.57\times10^{-3})\,\mathrm{SNR}^{1.80\pm0.06}$,
crossing the Jeffreys'~\cite{Jeffreys1961} substantial, strong, and decisive thresholds at
$\mathrm{SNR}=22.9$, $38.0$, and $55.9$ respectively. The same
near-quadratic form holds for the three remaining configurations
(Table~\ref{tab:summary}; full curves in Supplemental Material): reducing
the $\tilde\Lambda$ contrast to $\approx28\%$ (Curve~2, DD2$_{\epsilon_h}$ vs.\ a
near-degenerate DD2$_{\epsilon_l}$ hybrid) shifts every threshold to
higher SNR while leaving $n$ unchanged within uncertainty; swapping
which EOS is injected and which is recovered (Curve~3, SFHo injected,
DD2$_{\epsilon_h}$ recovered) reproduces Curve~1 to within $\sim20\%$ in threshold
SNR, showing the relation is not an artifact of a particular
true-EOS choice.

This exponent $n\approx1.8$--$2$ is not merely empirical. For the
zero-noise injections analyzed here, a Laplace/Occam-factor expansion
of the evidence integral gives, at leading order, a waveform mismatch
$\varepsilon$ between the correct- and wrong-EOS templates (minimized
over the wrong-EOS arm's nuisance parameters) that scales linearly
with the tidal-phase difference, so that
$\Delta\log Z \approx K\,\mathrm{SNR}^2$ with $K=\varepsilon^2/2$ and,
to leading tidal order, $\varepsilon \propto \Delta\tilde\Lambda$: a
quadratic scaling at leading order, with a contrast-dependent
prefactor $C \equiv K/\Delta\tilde\Lambda^{2}$ (see Supplemental
Material for the full derivation, including the noise-realization and
posterior-volume terms this leading approximation omits in general,
and the higher-order tidal corrections responsible for the residual
scatter in $C$). Fitting $C$ from Curves~1--3 gives
$C = (7.2\pm0.9)\times10^{-9}$, consistent across the three
independent contrast/role configurations; because $C$ also depends on
the binary's masses and spins, we treat this consistency as evidence
that the coefficient \emph{transfers} between configurations, not as
evidence that $C$ is universal. We used this calibration to make \emph{a priori} prediction: for a second binary mass point
($m_1=1.40\,M_\odot$, $m_2=1.20\,M_\odot$, $\Delta\tilde\Lambda
=518.9$) not used in the fit, the mean $C$ predicts a decisive-evidence
threshold of $\mathrm{SNR}\approx50.8$. We then ran the full
nested-sampling analysis for this mass point (Curve~4) and measured
$\mathrm{SNR}=50.7$ for decisive evidence [Fig.~\ref{fig:delta_logz_snr}(b)]
--- agreement well within the predicted uncertainty (central values differ by only 0.3\%),
and the empirical fit exponent $n=1.85\pm0.05$ for Curve~4 is
statistically indistinguishable from Curve~1. We emphasize that this 0.3\% test holds the detector network, waveform model, and EOS pair fixed relative to the calibration curves, isolating transferability across binary mass alone; transferability across EOS pair is tested separately, at a coarser $\sim30\%$
 level, in the spot-check campaign below (Discussion). The successful prediction of Curve 4 shows that the SNR threshold for GW-based EOS discrimination need not be determined solely through post hoc fitting, but can be estimated from a small number of calibration points.

\begin{table}[t]
    \caption{Power-law fit $\Delta\log Z = A\,\mathrm{SNR}^n$ and
    Jeffreys'-scale threshold SNRs for all four configurations. $q_1=\frac{1.232}{1.54}$; $q_2=\frac{1.2}{1.4}$.}
    \label{tab:summary}
    \begin{ruledtabular}
    \begin{tabular}{lccccc}
    Curve & $n$ & $A$ & sub. & strong & decisive \\
    \hline
    1: DD2$_{\epsilon_h}$$\to$SFHo, q$_1$          & 1.80 & $3.57\times10^{-3}$ & 22.9 & 38.0 & 55.9 \\
    2: DD2$_{\epsilon_h}$$\to$DD2$_{\epsilon_l}$,q$_1$   & 1.74 & $2.08\times10^{-3}$ & 35.1 & 59.4 & 88.6 \\
    3: SFHo$\to$DD2$_{\epsilon_h}$,q$_1$          & 1.95 & $1.46\times10^{-3}$ & 28.7 & 45.9 & 65.6 \\
    4: DD2$_{\epsilon_h}$$\to$SFHo, q$_2$          & 1.85 & $3.48\times10^{-3}$ & 21.3 & 34.9 & 50.7 \\
    \end{tabular}
    \end{ruledtabular}
\end{table}

\textit{Discussion.}---
The empirical exponents we measure ($n=1.74$--$1.95$) sit slightly
below, and trend toward, the asymptotic $n=2$ expected from a
Laplace/Occam-factor argument: at high SNR the leading deterministic
contribution to the log-evidence penalty for using the wrong template
scales as $\tfrac12\|\delta h_\perp\|^2 \propto \mathrm{SNR}^2$ for
the zero-noise injections analyzed here (Supplemental Material gives
the full derivation, including the noise-realization and
posterior-volume terms this leading term omits in general), the same
mismatch-based distinguishability criterion introduced by Lindblom,
Owen, and Brown for waveform-modeling accuracy
standards~\cite{Lindblom:2008cm}, and cast in the explicit
noise-weighted-inner-product form $\|\delta h\|^2 \simeq \langle
h_1|h_1\rangle + \langle h_2|h_2\rangle - 2\langle h_1|h_2\rangle_{\max}$
by Read~et al.\ in their study of numerical-relativity EOS
measurability~\cite{Read2013}. We flag one caveat on this exponent
explicitly: each curve's 8-point fit is leveraged heavily by its
single highest-SNR point, and the four-matched-point subset that
tightens the range to $n\approx1.90$--$1.99$ still includes that same
point, so this is not an anchor-independent confirmation. Dropping the
anchor entirely leaves only $\approx0.9$ decades of SNR per curve and
lowers the fitted exponent substantially, to $n\approx1.2$--$1.75$
(see the anchor-point sensitivity table in the Supplemental Material)
--- too narrow a range on its own
to pin down $n$ precisely. We therefore treat the approach of the
fitted exponent toward 2 as SNR increases, rather than a tightly
constrained $n\approx2$ independent of the anchor, as the more
defensible statement of this result. As a targeted check on this
leverage concern, we densified Curve~1 with three additional
zero-noise points at $\mathrm{SNR}=352$, $704$, and $1409$
($d_L=400$, $200$, $100\,$Mpc), filling the previously unprobed gap
between the eighth grid point and the anchor. The resulting 11-point
fit, $n=1.80\pm0.04$, is statistically unchanged from the original
8-point value, but the anchor-drop sensitivity shrinks substantially:
removing the highest-SNR point now lowers $n$ only to $1.75\pm0.05$,
compared with $1.46$ for the original 8-point grid, and the local
slope $K\equiv\Delta\log Z/\mathrm{SNR}^2$ decreases smoothly and
monotonically across all three new points with no curvature
(Supplemental Material). These results are consistent with the anchor-driven leverage arising primarily from the sparse sampling of the original SNR grid, rather than from a genuine discontinuity in the underlying relation.

Two systematic effects deserve discussion as caveats on $\Delta\log Z$
at the highest-SNR points. First, the wrong-EOS recovery arm's mass
ratio $q$ compensates for the EOS mismatch by drifting from its
injected value, and in Curve~3 (but not Curve~1) this drift is toward
the prior's $q_{\rm max}=1$ boundary, pinning against it at the
highest SNR and measurably inflating Curve~3's $\Delta\log Z$ relative
to Curve~1 at matched SNR. A related, distinct effect appears at
Curve~4's most extreme point (${\rm SNR}\approx3192$): the spin
($s_{1z}$) and coalescence-time ($t_c$) posteriors pin against their
own prior boundaries once the mass-ratio compensation channel is
exhausted, which may make the reported $\Delta\log Z$ there a
conservative lower bound. Neither caveat changes this work's
qualitative conclusions --- decisive distinguishability at extreme SNR
is established by a wide margin in every curve --- but both show that
prior-boundary choices deserve care in studies targeting the exact
high-SNR asymptote. Second, Curve~1 alone shows a $\sim3\sigma$ excess over Curve~3 confined to $\mathrm{SNR}\approx25$--$60$ whose origin we have not identified; its absence in Curve~4, which shares Curve~1's EOS-role assignment but not Curve~3's, tentatively suggests a role-assignment-specific feature rather than a generic pipeline artifact.

To probe the transferability of the calibrated coefficient $C$ beyond
the four systematic configurations studied above, we performed five
additional spot-checks spanning three further EOS pairs and two binary
mass points (Supplemental Material). These tests extend beyond the
relativistic mean-field (RMF) EOS used throughout the main analysis by
including the Skyrme functional SLy4~\cite{CHABANAT_1998231}. Four of
the five spot checks agree with the analytic prediction to within
$\sim15\%$, supporting the transferability of the calibrated scaling
beyond the original calibration set. The remaining case exhibits a
substantially smaller $\Delta\log Z$, coincident with the strongest
mass-ratio compensation observed in this study. The same EOS pair at a
different mass ratio shows no comparable suppression, suggesting that
unusually efficient parameter compensation, rather than the EOS pair
itself, limits the predictive accuracy of the simple scaling estimate
for particular binary configurations.

This study performs discrete Bayesian model selection between a
small, named set of candidate EOS: a decisive $\Delta\log Z$
establishes that the data prefer one specific template over another,
not that either is the true EOS of nature. This differs from the
flexible or non-parametric EOS inference (piecewise-polytrope,
spectral, or Gaussian-process EOS sampled jointly with the binary
parameters) used in most modern LVK analyses, where the true EOS can
in principle be recovered continuously without first supplying the
correct model from a finite list. Our framing is, however, directly
analogous to how LVK's own GW170817 analyses compare fixed tables of
nuclear-theory EOS, and the SNR thresholds reported here should be
read as conditional on the true EOS lying near one of the tested
candidates. This caveat is sharpened by the existence of ``tidal deformability doppelg\"angers''---pairs of physically distinct candidate EOS that differ substantially in pressure and radius at supranuclear densities but agree in $\tilde\Lambda(M)$ to $\Delta\tilde\Lambda\lesssim10$--$30$---for which our calibrated scaling implies decisive discrimination would require ${\rm SNR}\approx900$--$2800$, within reach of a 3G network only for the rare, GW170817-distance analogs it will observe, not for a typical detection~\cite{Raithel:2022efm,Raithel:2022aee}. Our results further assume zero detector noise (a single
representative realization at each SNR, not a noise-averaged
ensemble) and a fixed sky position and network geometry.
As a targeted check on this simplification, we reran Curve~1's
$d_L=2400\,$Mpc point with wide, uninformative priors on sky position
and inclination (uniform right ascension, and the standard isotropic
$\cos\delta$ and $\sin\iota$ priors on declination and inclination, in
place of the fixed values and narrow Gaussian used elsewhere) and
three independent real-noise realizations for each recovery model,
sharing an identical per-detector noise draw between the correct- and
wrong-EOS arms. This gives $\Delta\log Z=3.50\pm0.26$, a shift of
$1.09$ from the fixed-sky, zero-noise value at the same point
($4.59\pm0.37$) and well within the $\approx3.0$ scatter expected from
single-realization noise fluctuations alone (Supplemental Material)
--- evidence against, though not a full characterization of, an
artificial evidence inflation from these two simplifications acting
together. Extending this check to a full free-sky, noise-averaged campaign across every configuration in this study — multiple real-noise realizations at every SNR point, for both recovery arms, across all four curves and the spot-check draws — would multiply the nested-sampling cost of this study several fold and is left to future work.

\textit{Conclusion.}---
We have measured the Bayesian evidence for EOS discrimination in
simulated BNS mergers across four independent configurations spanning
tidal-deformability contrast, EOS true/wrong role, and binary mass,
and found that a single leading-order quadratic scaling law,
$\Delta\log Z \approx C\,(\tilde\Lambda_{\rm true}-\tilde\Lambda_{\rm wrong})^2\,\mathrm{SNR}^2$,
both fits the data well and, once its coefficient $C$ is calibrated on
three configurations alone, transfers predictively to a fourth's
decisive-evidence threshold to within $0.3\%$, tested before it was
measured. This relation converts an otherwise expensive full
nested-sampling campaign into a closed-form estimate of the SNR at
which a 3G network can decisively distinguish two candidate
neutron-star equations of state, providing a first-order design benchmark for ET and CE observational planning and for prioritizing
which BNS events are worth analyzing for EOS information. A five-draw
spot-check across three further EOS pairs and mass points shows this
transferability holds to within $\sim15\%$ in four of five cases. In such cases, the simple scaling estimate can fail by more than a factor of two when parameter compensation in the wrong-EOS recovery becomes unusually efficient (Discussion). We expect the quadratic scaling form itself --- a direct consequence of the Occam-factor argument (Discussion) --- to persist quite generally, while the coefficient C, which depends on the binary's masses, spins, and detector configuration, will require recalibration for each new EOS pair and mass regime rather than reuse across an ensemble.
Extending this validation to a broader ensemble of EOS pairs and mass
points, and relaxing the discrete-model-selection assumption (both
alongside the noise/sky-position robustness check already discussed
above), remain natural next steps toward a fully general predictive
framework for gravitational-wave EOS discrimination.

\textit{Code and data availability.}---
The nested-sampling pipeline (jimgw/BlackJAX NS-AW) used in this study is publicly available at ~\cite{jim_nested}. All the data related to the plot and tables in the main text and in the Supplemental Material is available in~\cite{Imam2026_ns}.

\begin{acknowledgments}
The author thanks Thibeau Wouters and Thomas C. K. Ng for helpful discussions on implementing the BlackJAX-accelerated nested-sampling (NS-AW) within the Jim framework, and Macarena Lagos for discussions on the 3G detector setup in the code.
\end{acknowledgments}
\bibliography{references}
\bibliographystyle{apsrev4-2}
\appendix
\setcounter{figure}{0}
\renewcommand{\thefigure}{S\arabic{figure}}

\setcounter{table}{0}
\renewcommand{\thetable}{S\arabic{table}}
\onecolumngrid
\begin{center}
{\Large\bfseries Supplemental Material}
\end{center}


This supplement provides (i) the full derivation of the analytic
scaling relation $\Delta\log Z \approx C(\dLt\cdot\rho)^2$ summarized
in the main text, (ii) the full per-point $\Delta \log Z$ data
underlying all four curves discussed there, (iii) the detailed
contrast-dependence and EOS-role-swap comparison figure referenced
there, (iv) the anchor-point leave-one-out sensitivity of each curve's
fitted power-law exponent $n$, (v) a targeted densification of
Curve~1 with three additional high-SNR points that directly probes
the anchor-leverage concern in (iv), and (vi) a free-sky,
free-inclination, real-noise-realization robustness check of the
zero-noise, fixed-sky simplification used throughout the main grid.

\begin{figure}
    \centering
    \includegraphics[width=0.5\textwidth]{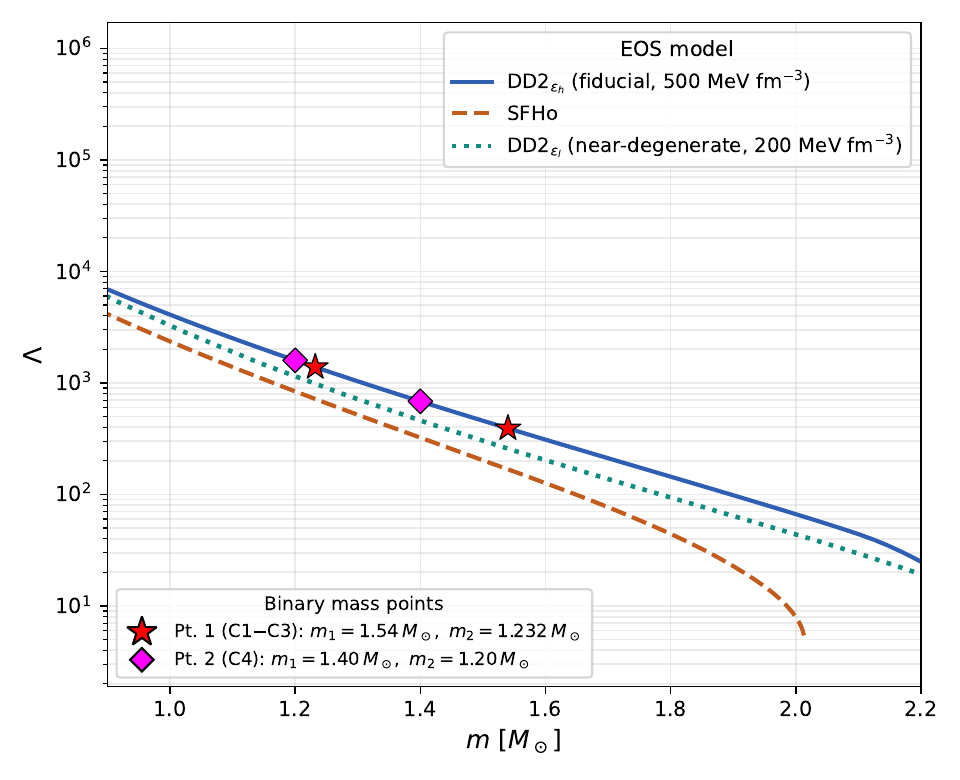}
    \caption{Mass--$\Lambda$ relation for the three EOS models used in
    this study. Stars mark mass point 1 ($m_1=1.54\,M_\odot$,
    $m_2=1.232\,M_\odot$), used in Curves~1--3; diamonds mark mass
    point 2 ($m_1=1.40\,M_\odot$, $m_2=1.20\,M_\odot$), used in
    Curve~4. DD2$_{\epsilon_l}$ is shown only at mass point 1, the only
    configuration in which it appears (Curve~2).}
    \label{fig:supp_mass_lambda}
\end{figure}

\textit{EOS models and binary mass points.}---Figure~\ref{fig:supp_mass_lambda}
shows the mass--$\Lambda$ relation for the three equations of state used
throughout this study (DD2$_{\epsilon_h}$, SFHo, and the near-degenerate
DD2$_{\epsilon_l}$), together with the two binary component-mass points
analyzed in the main text. SFHo is a relativistic
mean-field (RMF) hadronic model; DD2$_{\epsilon_h}$ and DD2$_{\epsilon_t}$ additionally
patches in a hybrid high-density segment (below), so Curve~2 already
probes hadronic-vs-hybrid, not only hadronic-vs-hadronic,
discrimination. These DD2$_{\epsilon_t}$ EOSs are constructed by
combining the DD2 EOS with a speed-of-sound ($c_s$) parameterized
high-density grid above the transition energy density
$\epsilon_t=\epsilon_h,\epsilon_l$: above $\epsilon_t$, $c_s^2/c^2$ is free to
vary between 0 and 1 across a 5-point grid, with $c_s^2$ matched
(continuous) to the DD2 value at $\epsilon_t$ itself, giving the
flexibility to generate an EOS with a desired $\tilde\Lambda$.

\textit{Detector network and sensitivity curves.}---The ET+2CE
network described in the main text is built with \texttt{jimgw}'s
\texttt{GroundBased2G} detector class: each Cosmic Explorer (CE)
instrument is an independent L-shaped interferometer, and the
Einstein Telescope (ET) is modeled as three co-located interferometers
with arms rotated 120\textdegree{} from one another, whose vertex
separations are propagated analytically via the spherical
forward-azimuth formula on a WGS-84 Earth model; each interferometer's
antenna pattern follows directly from its own detector tensor and
does not depend on this vertex-separation length. Strain sensitivity
curves are drawn from the public compilation distributed with
\texttt{gwfast}~\cite{IacovelliGWFAST2022}: CE uses the baseline
40\,km-arm strain amplitude spectral density (ASD) from the tunable
Cosmic Explorer design study~\cite{Srivastava2022CosmicExplorerDesign},
computed for a source $15\textdegree{}$ off normal incidence; ET uses
the 20\,km-arm, full high-frequency/low-frequency (HF/LF) cryogenic
``xylophone'' design power spectral density from the Einstein
Telescope design-comparison study~\cite{Branchesi2023ETDesigns}, whose
published file provides the separate HF and LF instrument PSDs
alongside their combined total, the last of which is the curve used
here. Each curve is loaded via \texttt{Detector.load\_and\_set\_psd}
(ASD values are squared internally to a PSD) and interpolated onto the
shared analysis frequency grid, $f_{\rm min}=5$~Hz to
$f_{\rm max}=2048$~Hz, before the heterodyned likelihood is evaluated.
\begin{figure}[tbp]
    \centering
    \includegraphics[width=0.5\textwidth]{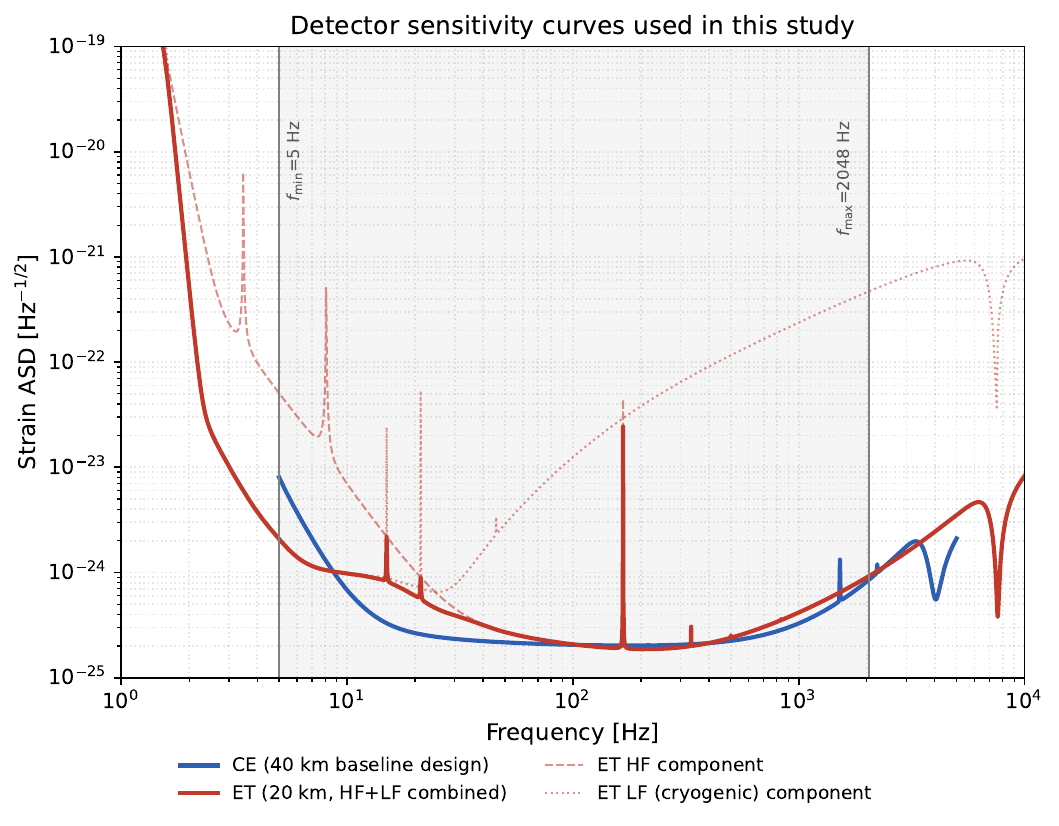}
    \caption{Strain amplitude spectral density (ASD) curves for the
    detector network used in this study. Cosmic Explorer (CE, navy)
    uses the baseline 40\,km-arm design~\cite{Srivastava2022CosmicExplorerDesign};
    Einstein Telescope (ET, crimson) uses the 20\,km-arm, full
    high-frequency/low-frequency (HF/LF) cryogenic ``xylophone''
    design~\cite{Branchesi2023ETDesigns}, shown both as its separate HF
    (dashed) and LF (dotted) instrument components and their combined
    total (solid) -- the curve actually used in the likelihood. Both
    curves are drawn from the public compilation distributed with
    \texttt{gwfast}~\cite{IacovelliGWFAST2022}. The shaded band marks
    the $f_{\rm min}=5$--$f_{\rm max}=2048$~Hz analysis range used
    throughout this study.}
    \label{fig:supp_sensitivity_curves}
\end{figure}

\begin{figure}[tbp]
\centering
\includegraphics[width=0.55\textwidth]{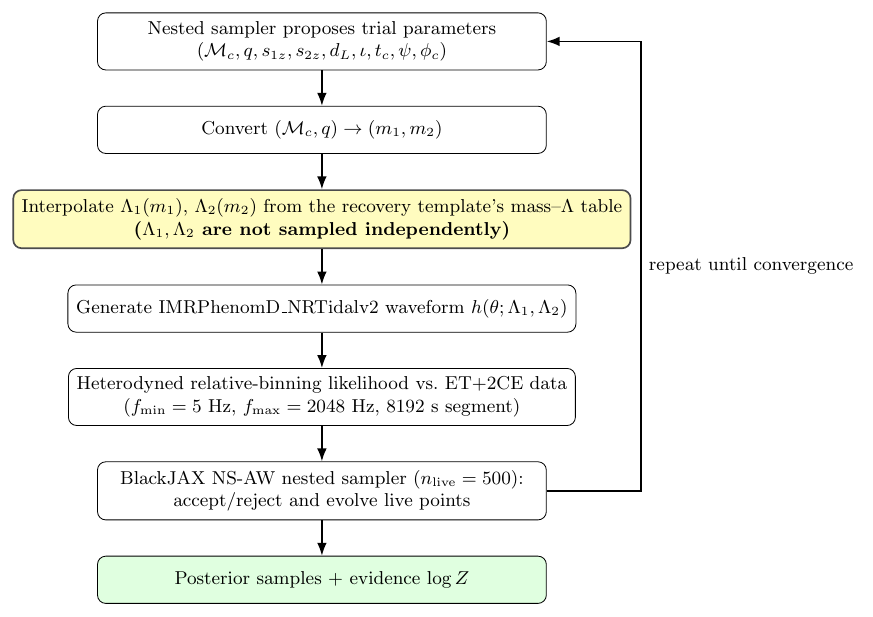}
\caption{Bayesian inference pipeline used for every recovery run.
Each nested-sampling trial's chirp mass and mass ratio are converted
to component masses, which are then used to look up (not sample)
$\Lambda_1$, $\Lambda_2$ from the recovery template's own EOS
mass--$\Lambda$ table (highlighted) before the waveform and
likelihood are evaluated; comparing a run whose table matches the
injected EOS against one whose table does not is what generates the
$\Delta\log Z$ statistic reported throughout this work.}
\label{fig:supp_bayes_flowchart}
\end{figure}

\textit{Bayesian inference setup.}---As summarized in the main-text
Method paragraph, every recovery run in this study uses the
\texttt{jimgw} nested-sampling package~\cite{Edwards:2023sak,Wong:2023lgb,Wouters:2024oxj}
with a BlackJAX-accelerated nested-sampling annealed-walkers (NS-AW)
sampler ($n_{\rm live}=500$), a heterodyned relative-binning
likelihood, and the IMRPhenomD\_NRTidalv2~\cite{Dietrich_2017,Dietrich_2019}
waveform, on an ET+2CE detector network with $f_{\rm min}=5\,$Hz,
$f_{\rm max}=2048\,$Hz, and an $8192\,$s analysis segment. Sampled
parameters are $(\mathcal{M}_c,q,s_{1z},s_{2z},d_L,\iota,t_c,\psi,\phi_c)$:
chirp mass and mass ratio use narrow uniform priors bracketing the
injected value at each mass point; spins use
$s_{1z},s_{2z}\sim\mathrm{Uniform}(-0.05,0.05)$; luminosity distance
uses a power-law prior ($\alpha=2$) at seven of the eight SNR grid
points, replaced by a narrow Gaussian anchored at the injected value
for the highest-SNR anchor point (main text)~\cite{Abbott2017MMA}. Together with this
distance prior, inclination uses a narrow Gaussian
($\mu=2.8495$, $\sigma=0.03$) rad\citep{Hotokezaka2019Hubble} centered on its GW170817-informed value,
and sky location at the injected
value~\cite{Abbott2017MMA} across all main-text runs (main text); these three choices
jointly replicate a GW170817-``like'' event with
electromagnetic-counterpart-informed localization at the anchor point.
A point easily missed: $\Lambda_1$ and $\Lambda_2$ are not themselves
free sampled parameters. At every likelihood evaluation the sampler's
trial $(\mathcal{M}_c,q)$ is converted to component masses
$(m_1,m_2)$, and $\Lambda_1(m_1)$, $\Lambda_2(m_2)$ are read off the
recovery template's own mass--$\Lambda$ table (DD2$_{\epsilon_h}$, SFHo, or
DD2$_{\epsilon_l}$, depending on which arm) by linear interpolation ---
the same mechanism (\texttt{EOSMcEtaToLambdaTransform}) used
throughout this study to distinguish a ``correct-EOS'' from a
``wrong-EOS'' recovery run. Figure~\ref{fig:supp_bayes_flowchart}
summarizes this pipeline.
 
\textit{Derivation of the analytic scaling relation.}---
We derive the leading-order relation between $\Delta\log Z$ and the
tidal-deformability contrast used in the main text, and are explicit
about every approximation involved, since this relation is used to
make a genuine out-of-sample prediction (below) rather than fit only
after the fact.

Consider two hypotheses $H_{\rm inj}$ and $H_{\rm alt}$: $H_{\rm inj}$'s
template EOS matches the injected model, $H_{\rm alt}$'s does not, and
the two share identical functional form, parameter dimensionality, and
priors except for the fixed tidal deformabilities $(\Lambda_1,\Lambda_2)$
each assumes. (Following the main text, we call $H_{\rm inj}$ ``true" or
``correct" and $H_{\rm alt}$ ``wrong" as shorthand; per the scope caveat
discussed there, this labels the injected model relative to this
simulated study only, not a claim about the true EOS of nature.) Under
a Laplace approximation, each hypothesis' evidence expands about its
best fit $\hat\theta$ as $\log Z_M \simeq \log\mathcal L(\hat\theta_M) +
\log\pi(\hat\theta_M) + \tfrac12\log|2\pi\Sigma_M|$, so
$\Delta\log Z = \Delta\log\mathcal L + \Delta_{\rm vol}$ with
$\Delta_{\rm vol}\equiv\Delta\log\pi+\tfrac12\log(|\Sigma_{\rm inj}|/
|\Sigma_{\rm alt}|)$. Because $H_{\rm inj}$ and $H_{\rm alt}$ share
identical dimensionality and prior ranges, the explicit dimensional
Occam penalty cancels between them; it does \emph{not} follow that
$\Delta_{\rm vol}$ vanishes entirely, since the local posterior
covariances $\Sigma_{\rm inj}\neq\Sigma_{\rm alt}$ can differ once the
wrong-EOS arm's posterior compensates in mass ratio, spin, and
coalescence time (as documented empirically in the main text's
wall-pinning discussion). We do not compute $\Delta_{\rm vol}$
analytically; we instead treat it, along with higher-order tidal
terms (below), as a subleading contribution absorbed into the
$\sim30\%$ curve-to-curve scatter in the calibrated prefactor $C$.

For the likelihood term, writing the Gaussian likelihood
$\ln p(d\mid\theta,H) = -\tfrac12\langle d-h(\theta)\mid
d-h(\theta)\rangle + \text{const}$ with injected data
$d=h_{\rm inj}(\theta_0)+n$ and $\hat\theta_{\rm inj}\approx\theta_0$
(the correct-EOS fit recovers the truth),  using
$d-h_{\rm inj}=n$ and $d-h_{\rm alt}(\hat\theta_{\rm alt})=n+\delta
h_\perp$ gives, for a single noise realization,
\begin{equation}
  \Delta\log\mathcal L \;=\; \langle n\mid\delta h_\perp\rangle \;+\;
  \tfrac12\,\lVert\delta h_\perp\rVert^2, \qquad
  \delta h_\perp \equiv h_{\rm inj}(\theta_0) - h_{\rm alt}(\hat\theta_{\rm alt}),
  \label{eq:supp_step1}
\end{equation}
where the $\perp$ subscript records that $\hat\theta_{\rm alt}$ has
already been optimized over every nuisance parameter, so $\delta
h_\perp = P_\perp\,\delta h_{\rm EOS}$ is the projection of the raw
EOS-induced waveform difference orthogonal to the tangent space of
those nuisance directions --- the object actually responsible for the
mass-ratio, spin, and coalescence-time compensation, and the
associated prior-boundary wall-pinning, reported throughout the main
text. $\lVert\delta h_\perp\rVert\equiv\sqrt{\langle\delta h_\perp\mid
\delta h_\perp\rangle}$ is the same noise-weighted
mismatch/distinguishability statistic underlying the waveform-modeling
accuracy standard of Lindblom, Owen \& Brown~\cite{Lindblom:2008cm}
and cast in this explicit noise-weighted-inner-product form by Read~et
al.~\cite{Read2013}.

The noise cross-term $\langle n\mid\delta h_\perp\rangle$ in
Eq.~\eqref{eq:supp_step1} averages to zero over noise realizations, so
in general only the noise-averaged statement $\langle\Delta\log
Z\rangle_n\simeq\tfrac12\lVert\delta h_\perp\rVert^2+\langle
\Delta_{\rm vol}\rangle_n$ holds; for a single noisy realization,
$\Delta\log\mathcal L$ can depart from $\tfrac12\lVert\delta
h_\perp\rVert^2$ by a term of order $\lVert\delta h_\perp\rVert$
itself. Every injection in this study, however, uses \emph{zero}
detector noise --- $n=0$ exactly, not a noise average --- so the cross
term vanishes identically for the specific realizations analyzed here,
and dropping the subleading $\Delta_{\rm vol}$ term,
\begin{equation}
  \Delta\log Z \;\approx\; \tfrac12\,\lVert\delta h_\perp\rVert^2
  \label{eq:supp_step1b}
\end{equation}
holds for our simulations as an exact consequence of the zero-noise
injection, not as a generic identity for $\Delta\log Z$ at arbitrary
noise realizations.

To make the SNR-scaling explicit without conflating the network
amplitude with the waveform shape, write the template as $\tilde
h(f)=\rho\,\hat h_0(f)$, where $\hat h_0$ is a unit-SNR waveform shape
($\langle\hat h_0\mid\hat h_0\rangle=1$) so that $\rho^2=\langle\tilde
h\mid\tilde h\rangle$ is, by this definition, the network SNR; the
true and wrong-EOS templates share this amplitude and differ, to the
order relevant here, only in phase,
$\tilde h_{\rm inj/alt}(f)=\rho\,\hat h_0(f)\,e^{i\Psi_{\rm inj/alt}(f)}$.
For a small phase difference $\delta\Psi_\perp\equiv[\Psi_{\rm inj}-
\Psi_{\rm alt}]_\perp$ (projected as above), $\delta h_\perp(f)\approx
i\rho\,\hat h_0(f)\,\delta\Psi_\perp(f)$, so
\begin{equation}
  \lVert\delta h_\perp\rVert^2 \;\approx\; \rho^2\times
  \underbrace{4\!\int_0^\infty\!\frac{|\hat h_0(f)|^2\,[\delta\Psi_\perp(f)]^2}{S_n(f)}\,df}_{\displaystyle\equiv\,J},
  \label{eq:supp_step2}
\end{equation}
with $J$ manifestly independent of distance because $\hat h_0$ is
unit-normalized: $|\tilde h(f)|^2$ itself scales as $1/d_L^2$, so only
the unit-normalized $\hat h_0$ -- not $\tilde h$ itself -- gives a
genuinely distance-independent $J$.

At leading post-Newtonian order, the tidal phase in this study's
\texttt{IMRPhenomD\_NRTidalv2} templates depends linearly on
$\tilde\Lambda$, $\Psi_{\rm tidal}(f;m_1,m_2,\tilde\Lambda)\approx
\tilde\Lambda\,g_1(m_1,m_2,f)$; the full waveform model contains
additional higher-order tidal terms that make this only a
leading-order statement, not an exact one, so
$\delta\Psi_\perp(f)\approx\dLt\,g_1(f)$ with
$\dLt\equiv\tilde\Lambda_{\rm inj}-\tilde\Lambda_{\rm alt}$ up to
corrections we do not track analytically. Substituting into
Eq.~\eqref{eq:supp_step2} and factoring $(\dLt)^2$ out of the integral,
$J\approx(\dLt)^2\,I_\perp(m_1,m_2,\text{network})$ with $I_\perp\equiv
4\int_0^\infty|\hat h_0(f)|^2g_1(f)^2/S_n(f)\,df$, giving the combined
leading-order scaling law
\begin{equation}
  \boxed{\;\Delta\log Z \;\approx\; C\,\big(\dLt\cdot\rho\big)^2\;},
  \qquad C\equiv I_\perp/2,
  \label{eq:supp_master}
\end{equation}
in which tidal contrast and loudness enter only through their product.
Both the linear tidal-phase approximation and the neglected
$\Delta_{\rm vol}$ term are leading-order truncations, and we regard
Eq.~\eqref{eq:supp_master} as the leading term in an expansion, not an
exact statement --- consistent with the $\sim30\%$ curve-to-curve
scatter in the calibrated $C$ (below) and the anchor-point sensitivity
of the fitted exponent $n$ (Table~\ref{tab:anchor}), rather than an
indication of a genuine departure from quadratic scaling.

Because $C\equiv I_\perp/2$ depends on the binary's masses, spins,
sky position and orientation, the detector network, the frequency
range analyzed, the waveform model, and (through $\Delta_{\rm vol}$)
the nuisance-parameter priors, it is not a universal physical
constant. The Curve~4 test below is therefore best read as an
empirical test of whether a coefficient calibrated at one binary
configuration transfers predictively to another with a different
chirp mass and mass ratio, not as evidence that $C$ is
mass-independent in general.

Inverting Eq.~\eqref{eq:supp_master} gives the \emph{analytic}
threshold-crossing SNR for a conventional Jeffreys'-scale label $T$
(substantial $=1$, strong $=2.5$, decisive $=5$ --- labels of
convenience on the continuous curve $\Delta\log Z(\rho)$, not
fundamental statistical thresholds): $\rho_T=(1/\dLt)\sqrt{T/C}$. The
\emph{empirical} threshold SNRs quoted in the main text and
Table~\ref{tab:anchor} are instead obtained by inverting each curve's
fitted power law, $\rho_T=(T/A)^{1/n}$, using the 8-point fit unless
stated otherwise; we use this fit-based inversion, rather than linear
interpolation between adjacent data points, throughout.

The prefactor $C$ is measured from the highest-SNR (``anchor") point of
each curve ($d_L=42\,$Mpc, $\rho\approx3355$), where the zero-noise
simplification behind Eq.~\eqref{eq:supp_step1b} is exact and the
high-SNR Laplace approximation is most accurate, via
$K\equiv\Delta\log Z/\rho^2$; then $C=K/(\dLt)^2$ using each curve's
tidal-deformability contrast at mass point 1
($\tilde\Lambda_{\rm DD2_{\epsilon_h}}=743.01$, $\tilde\Lambda_{\rm SFHo}=361.05$,
$\tilde\Lambda_{\rm DD2_{\epsilon_l}}=532.32$), summarized in
Table~\ref{tab:C_calibration}.

\renewcommand{\arraystretch}{0.5}
\begin{table*}
\caption{Calibration of the scaling prefactor $C$ from the anchor
point ($d_L=42\,$Mpc, $\rho\approx3355$) of Curves 1--3 (mass point
1). The mean $\bar C$ is used, together with mass point 2's
tidal-deformability contrast, to make the pre-registered Curve~4
prediction in Table~\ref{tab:curve4_prediction}.}
\label{tab:C_calibration}
\begin{ruledtabular}
\begin{tabular}{lcccc}
Curve & $\Delta\log Z$ at anchor & $K=\Delta\log Z/\rho^2$ & $\dLt$ & $C=K/(\dLt)^2$ \\
\hline
1 (DD2$_{\epsilon_h}$-true/SFHo-wrong)             & 10142.31 & $9.01\times10^{-4}$ & 381.96 & $6.18\times10^{-9}$ \\
2 (DD2$_{\epsilon_h}$-true/DD2$_{\epsilon_l}$-wrong) & 4003.09  & $3.56\times10^{-4}$ & 210.69 & $8.01\times10^{-9}$ \\
3 (SFHo-true/DD2$_{\epsilon_h}$-wrong, role-swap)  & 12093.76 & $1.07\times10^{-3}$ & 381.96 & $7.37\times10^{-9}$ \\
\hline
\multicolumn{4}{r}{Mean $\bar C$ (Curves 1--3)} & $7.19\times10^{-9}$ \\
\end{tabular}
\end{ruledtabular}
\end{table*}

$C$ clusters within a $\sim$30\% band across two very different tidal
contrasts (51.4\% and 28.4\%) and both EOS-role assignments, supporting
the $(\dLt)^2$ scaling as a leading-order description; the residual
spread is attributed to higher-PN tidal terms neglected in the phase
expansion above, mass-ratio marginalization effects, and the
prior-boundary wall-pinning systematic discussed in the main text.

This calibration, performed entirely from Curves 1--3 (mass point 1),
enables a genuine test of whether $C$ \emph{transfers} predictively to
an independent binary configuration, rather than a demonstration that
$C$ is universal. Using $\bar C=7.19\times10^{-9}$ and mass point 2's contrast
($\dLt_4=518.91$) --- fixed before any Curve 4 data existed ---
Eq.~\eqref{eq:supp_master} predicts threshold SNRs of 22.7
(substantial), 36.0 (strong), and 50.8 (decisive). Propagating the
$\pm12.9\%$ calibration scatter in $\bar C$ (Curves 1--3) through
$\rho_T\propto C^{-1/2}$ via Monte Carlo sampling gives 95\%
prediction intervals of $[20.3,26.3]$, $[32.1,41.6]$, and
$[45.4,58.8]$ respectively. The subsequently completed Curve 4
campaign measured these thresholds empirically (via the fit-based
inversion described above) at 21.3, 34.9, and 50.7 respectively
(Table~\ref{tab:anchor} and main-text Table~I) --- each falling
within its pre-registered 95\% interval, with agreement to the
central prediction improving monotonically toward the highest-SNR
threshold ($7\%$, $3\%$, and $0.3\%$ respectively) exactly as
expected for a high-SNR asymptotic approximation. We regard this
transferability, not the fitted exponent alone, as the paper's
central quantitative result. Table~\ref{tab:curve4_prediction}
summarizes this comparison.

\begin{table}[h]
\caption{Curve~4 out-of-sample prediction test: threshold SNRs
predicted from $\bar C$ (Table~\ref{tab:C_calibration}) and mass
point 2's tidal contrast, fixed before any Curve~4 data existed;
the 95\% prediction interval propagated from the $\pm12.9\%$
calibration scatter in $\bar C$ via Monte Carlo sampling of
$\rho_T\propto C^{-1/2}$; and the subsequently measured empirical
threshold (Table~\ref{tab:anchor} and main-text Table~I). All three
measured values fall within their pre-registered interval.}
\label{tab:curve4_prediction}
\begin{ruledtabular}
\begin{tabular}{lcccc}
Threshold & Predicted SNR & 95\% interval & Measured SNR & In interval \\
\hline
Substantial & 22.7 & [20.3, 26.3] & 21.3 & Yes \\
Strong      & 36.0 & [32.1, 41.6] & 34.9 & Yes \\
Decisive    & 50.8 & [45.4, 58.8] & 50.7 & Yes \\
\end{tabular}
\end{ruledtabular}
\end{table}

\begin{figure}[tbp]
    \centering
    \includegraphics[width=0.95\textwidth]{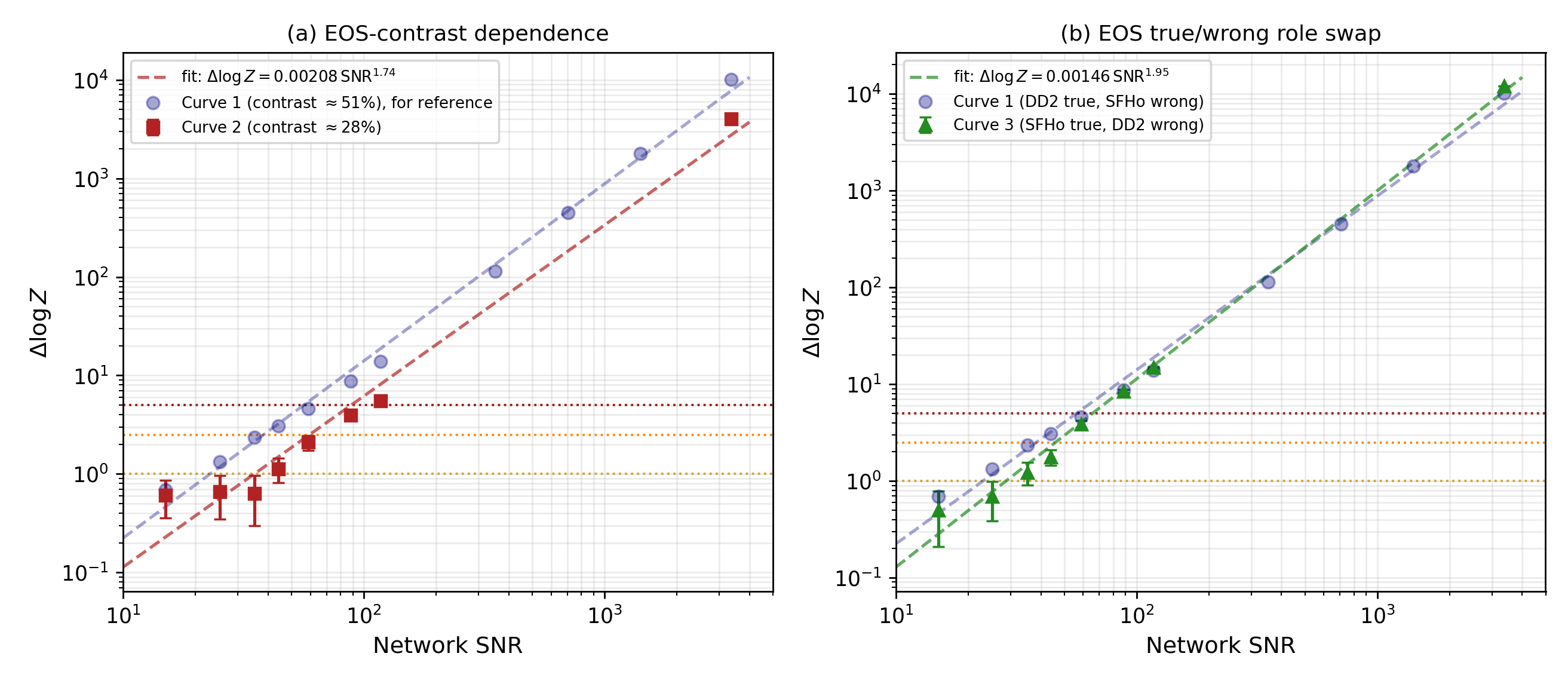}
    \caption{(a) Curve~2: $\Delta \log Z$ vs.\ network SNR for
    DD2$_{\epsilon_h}$(true) vs.\ DD2$_{\epsilon_l}$(wrong), a near-degenerate EOS
    pair with $\tilde\Lambda$ contrast $\approx28\%$, compared against
    Curve~1's $\approx51\%$-contrast result (faint navy). Reducing
    contrast shifts every Jeffreys' threshold to higher SNR while
    leaving the fitted exponent $n$ unchanged within uncertainty. (b)
    Curve~3: the same comparison for SFHo(true) vs.\ DD2$_{\epsilon_h}$(wrong) --- an
    EOS true/wrong role swap relative to Curve~1 at otherwise identical
    injected parameters. The two curves agree to within their
    statistical uncertainty over most of the SNR range; the
    $\mathrm{SNR}\approx25$--$60$ deviation is discussed in the main
    text.}
    \label{fig:supp_curves23}
\end{figure}

Figure~\ref{fig:supp_curves23} shows
Curve~2 (EOS-contrast dependence) and Curve~3 (EOS true/wrong role
swap) in full, each plotted against a faint reproduction of Curve~1
for reference; both are summarized only by their fitted $(n,A)$ and
threshold SNRs in the main-text table.

\begin{figure}[tbp]
    \centering
    \includegraphics[width=0.95\textwidth]{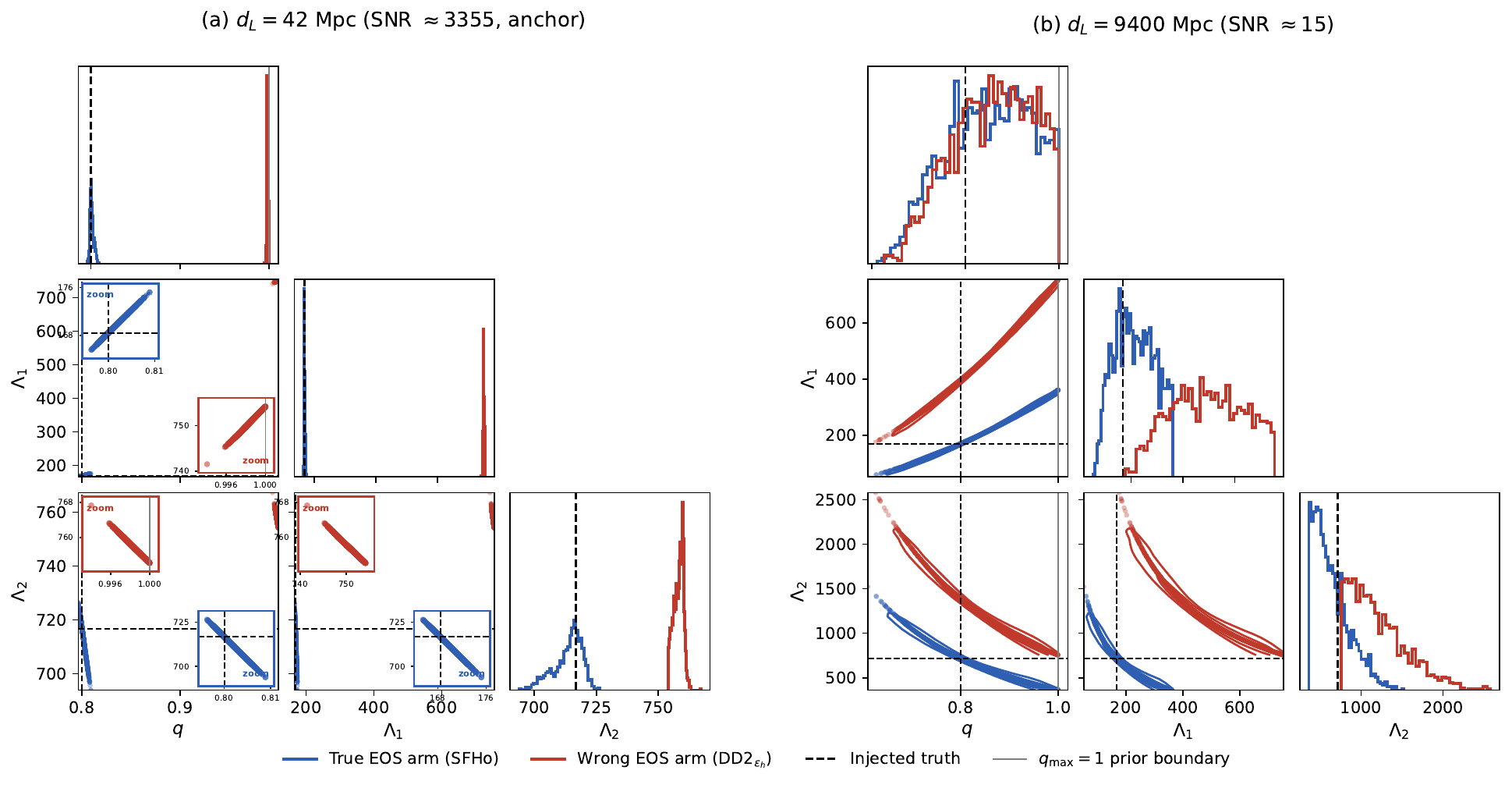}
    \caption{Overlaid posteriors in $(q,\Lambda_1,\Lambda_2)$ for
    Curve~3's true-EOS (SFHo, navy) and wrong-EOS (DD2$_{\epsilon_h}$, crimson)
    recovery arms, with the injected truth (dashed) and the $q_{\max}=1$
    prior boundary (solid gray) marked. (a) At the $d_L=42$ anchor
    (SNR$\,\approx\,$3355), the true arm recovers $q=0.801\pm0.002$
    (unbiased), while the wrong arm is pinned against the $q_{\max}=1$
    boundary at $q=0.998\pm0.001$. (b) At $d_L=9400$
    (SNR$\,\approx\,$15), both arms recover consistent, unpinned
    posteriors ($q=0.853\pm0.089$ and $q=0.864\pm0.084$
    respectively), confirming the pinning is an SNR-driven effect, not
    a prior artifact present at every SNR.}
    \label{fig:supp_wallpinning}
\end{figure}

\textit{Wall-pinning corner plot.}---Figure~\ref{fig:supp_wallpinning}
shows the mechanism behind Curve~3's prior-boundary pinning directly in
posterior-sample space: overlaid $(q,\Lambda_1,\Lambda_2)$ corner plots
for the true-EOS (SFHo) and wrong-EOS (DD2$_{\epsilon_h}$) recovery arms, at the
high-SNR anchor and the lowest-SNR grid point.

\textit{Full per-point data.}---Tables~\ref{tab:curve1}--\ref{tab:curve4}
give every $(\mathrm{SNR}, d_L, \Delta\log Z)$ triple used in the
main-text fits, together with the $1\sigma$ nested-sampling evidence
uncertainty reported by the sampler at each point.

\begin{table}[h]
\caption{Curve 1: DD2$_{\epsilon_h}$(true) vs.\ SFHo (wrong), mass point 1
($m_1=1.54\,M_\odot$, $m_2=1.232\,M_\odot$), $\tilde\Lambda$ contrast
$\approx51.4\%$. The three points at $d_L=400,200,100\,$Mpc (marked
$^\dagger$) are the densification points discussed below, run under
the same zero-noise, fixed-sky protocol as the original eight points.}
\label{tab:curve1}
\begin{ruledtabular}
\begin{tabular}{ccc}
SNR & $d_L$ (Mpc) & $\Delta \log Z$ \\
\hline
14.99 & 9400 & $0.69 \pm 0.26$ \\
25.16 & 5600 & $1.33 \pm 0.29$ \\
35.22 & 4000 & $2.34 \pm 0.31$ \\
44.03 & 3200 & $3.07 \pm 0.31$ \\
58.70 & 2400 & $4.59 \pm 0.37$ \\
88.06 & 1600 & $8.71 \pm 0.32$ \\
117.41 & 1200 & $13.85 \pm 0.38$ \\
352.24$^\dagger$ & 400 & $114.44 \pm 0.42$ \\
704.49$^\dagger$ & 200 & $450.69 \pm 0.43$ \\
1408.97$^\dagger$ & 100 & $1792.42 \pm 0.42$ \\
3354.7 & 42 & $10142.31 \pm 0.47$ \\
\end{tabular}
\end{ruledtabular}
\end{table}

\begin{table}[h]
\caption{Curve 2: DD2$_{\epsilon_h}$ (true) vs.\ DD2$_{\epsilon_l}$ (wrong,
near-degenerate), mass point 1, $\tilde\Lambda$ contrast $\approx28.4\%$.}
\label{tab:curve2}
\begin{ruledtabular}
\begin{tabular}{ccc}
SNR & $d_L$ (Mpc) & $\Delta \log Z$ \\
\hline
14.99 & 9400 & $0.61 \pm 0.25$ \\
25.16 & 5600 & $0.66 \pm 0.31$ \\
35.22 & 4000 & $0.63 \pm 0.33$ \\
44.03 & 3200 & $1.13 \pm 0.32$ \\
58.70 & 2400 & $2.10 \pm 0.37$ \\
88.06 & 1600 & $3.94 \pm 0.33$ \\
117.41 & 1200 & $5.52 \pm 0.38$ \\
3354.7 & 42 & $4003.09 \pm 0.45$ \\
\end{tabular}
\end{ruledtabular}
\end{table}

\begin{table}[h]
\caption{Curve 3: SFHo (true) vs.\ DD2$_{\epsilon_h}$ (wrong), EOS-role swap relative
to Curve 1, mass point 1.}
\label{tab:curve3}
\begin{ruledtabular}
\begin{tabular}{ccc}
SNR & $d_L$ (Mpc) & $\Delta \log Z$ \\
\hline
14.990 & 9400 & $0.50 \pm 0.29$ \\
25.162 & 5600 & $0.69 \pm 0.30$ \\
35.226 & 4000 & $1.23 \pm 0.32$ \\
44.033 & 3200 & $1.77 \pm 0.33$ \\
58.71 & 2400 & $3.86 \pm 0.32$ \\
88.07 & 1600 & $8.46 \pm 0.36$ \\
117.42 & 1200 & $14.83 \pm 0.40$ \\
3354.88 & 42 & $12093.76 \pm 0.45$ \\
\end{tabular}
\end{ruledtabular}
\end{table}

\begin{table}[h]
\caption{Curve 4: DD2$_{\epsilon_h}$ (true) vs.\ SFHo (wrong), mass point 2
($m_1=1.40\,M_\odot$, $m_2=1.20\,M_\odot$), the mass-generalization
test and out-of-sample analytic-prediction validation.}
\label{tab:curve4}
\begin{ruledtabular}
\begin{tabular}{ccc}
SNR & $d_L$ (Mpc) & $\Delta \log Z$ \\
\hline
14.2641 & 9400 & $0.64 \pm 0.28$ \\
23.9432 & 5600 & $1.29 \pm 0.31$ \\
33.5206 & 4000 & $2.09 \pm 0.32$ \\
41.9007 & 3200 & $3.83 \pm 0.30$ \\
55.8676 & 2400 & $5.26 \pm 0.38$ \\
83.8013 & 1600 & $10.91 \pm 0.34$ \\
111.7351 & 1200 & $18.14 \pm 0.33$ \\
3192.4327 & 42 & $12359.66 \pm 0.47$ \\
\end{tabular}
\end{ruledtabular}
\end{table}

\textit{Curve 1 densification and anchor-leverage robustness
check.}---The anchor-point sensitivity documented below (originally:
dropping each curve's single highest-SNR point lowers its fitted
exponent substantially, since the remaining seven points span under
one decade in SNR) raises an obvious question: is the anchor simply
extending an otherwise well-sampled power law, or is it pulling the
fit toward a value the data do not otherwise support? The eight-point
grid used for all four curves has no points between $\mathrm{SNR}
\approx117$ (Curve 1's $d_L=1200$ point) and the anchor at
$\mathrm{SNR}\approx3355$ --- a gap of $1.46$ decades with zero
coverage. We closed this gap for Curve~1, the fiducial and most
extensively cross-validated configuration, with three additional
zero-noise points at $d_L=400,200,100\,$Mpc
($\mathrm{SNR}=352.24,704.49,1408.97$), using the same fixed-sky,
zero-noise configuration and the same injected binary as the original
eight points, with power-law $d_L$ priors widened to
$[150,1200]$, $[70,600]$, and $[30,300]\,$Mpc respectively to
accommodate the higher injected SNR at each point. We did not repeat
this densification for Curves~2--4, both for computational cost and
because, as shown below, the result is consistent with a generic
feature of the pipeline rather than something specific to Curve~1;
Table~\ref{tab:anchor} already shows all four curves collapse to a
similar reduced exponent when their own anchor is dropped, which is
the pattern this check is designed to explain.

Table~\ref{tab:curve1_K} gives the local slope $K\equiv\Delta\log
Z/\mathrm{SNR}^2$ at every point from $d_L=1200$ to the anchor. $K$
decreases smoothly and monotonically across the three new points, with
no jump or sign of curvature at any of them --- the three densification
points interpolate cleanly between the $d_L=1200$ value and the
anchor's asymptotic value, rather than revealing hidden structure in
the previously unprobed gap.

\begin{table}[h]
\caption{Local slope $K=\Delta\log Z/\mathrm{SNR}^2$ across the
densified region of Curve~1, from the last original grid point through
the anchor. The three densification points (marked $^\dagger$) fall
smoothly between their neighbors with no curvature.}
\label{tab:curve1_K}
\begin{ruledtabular}
\begin{tabular}{ccc}
SNR & $d_L$ (Mpc) & $K=\Delta\log Z/\mathrm{SNR}^2$ \\
\hline
117.41 & 1200 & $1.0047\times10^{-3}$ \\
352.24$^\dagger$ & 400 & $9.223\times10^{-4}$ \\
704.49$^\dagger$ & 200 & $9.081\times10^{-4}$ \\
1408.97$^\dagger$ & 100 & $9.029\times10^{-4}$ \\
3354.70 & 42 (anchor) & $9.012\times10^{-4}$ \\
\end{tabular}
\end{ruledtabular}
\end{table}

\begin{figure}[h]
    \centering
    \includegraphics[width=0.95\textwidth]{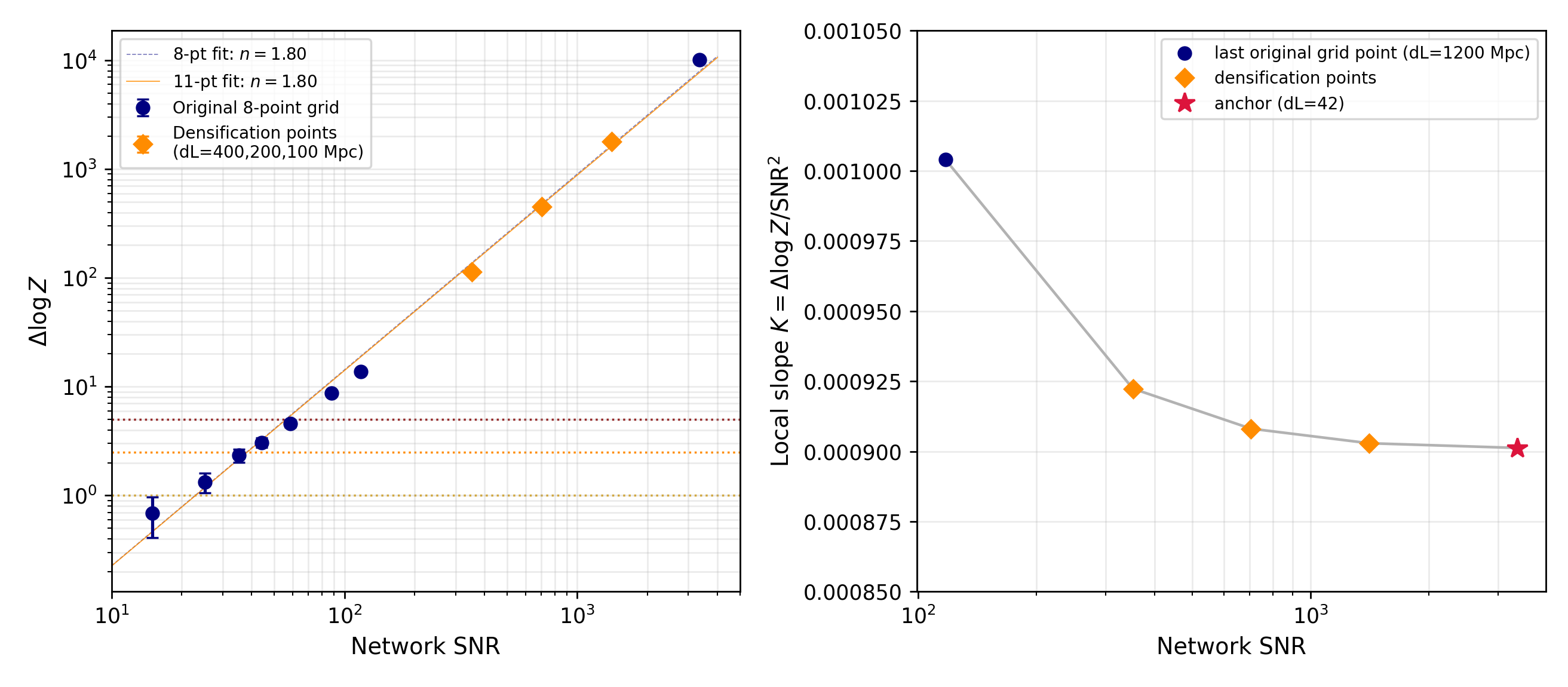}
    \caption{(a) Curve~1's full densified grid: the original 8 points
    (navy circles) and the 3 new $d_L=400,200,100\,$Mpc points (orange
    diamonds), with the original 8-point fit and the new 11-point fit
    overlaid (the two are statistically indistinguishable). (b) The
    local slope $K=\Delta\log Z/\mathrm{SNR}^2$ across the previously
    unprobed gap between the last original grid point ($d_L=1200$) and
    the anchor ($d_L=42$): the three densification points interpolate
    smoothly between the two, with no curvature or discontinuity.}
    \label{fig:supp_curve1_densification}
\end{figure}

Figure~\ref{fig:supp_curve1_densification} shows this visually: panel
(a) plots the full 11-point grid, with the three densification points
distinguished from the original eight, alongside both the original
8-point fit and the new 11-point fit;
panel (b) plots the local $K$ values from Table~\ref{tab:curve1_K}
directly, showing the smooth, monotonic, curvature-free decline from
the last original grid point through the anchor.

Refitting Curve~1's power law with all 11 points gives
$n=1.7973\pm0.0411$, $A=3.573\times10^{-3}$, statistically
unchanged from the original 8-point fit ($n=1.8008\pm0.0610$) and
shifting the Jeffreys' threshold SNRs by under $1\%$ (substantial:
$22.9\to23.0$; strong: $38.0\to38.3$; decisive: $55.9\to56.3$). The
leave-one-out anchor sensitivity, however, improves substantially:
dropping the $d_L=42$ anchor from the 11-point grid lowers $n$ only to
$1.7505\pm0.0467$ (a $0.047$ shift), compared with the collapse from
$1.80$ to $1.46$ (a $0.34$ shift) seen when the anchor is dropped from
the original, undensified 8-point grid --- see the added row in
Table~\ref{tab:anchor}. Three additional points, filling what had been
the single largest gap in the SNR coverage, are sufficient to reduce
the anchor's leverage on the fitted exponent by roughly a factor of
seven. We read this, together with the clean $K$-continuity in
Table~\ref{tab:curve1_K}, as evidence that the anchor-driven exponent
sensitivity reported below is a consequence of sparse high-SNR
sampling in the original grid rather than a genuine break in the
underlying $\Delta\log Z(\mathrm{SNR})$ relation --- supporting the
main text's interpretation of the fitted exponent as trending smoothly
toward the asymptotic Occam-factor value $n\to2$, rather than being
artificially inflated by a single unrepresentative point.

\textit{Anchor-point sensitivity.}---Because each curve's fit is
dominated in leverage by its single highest-SNR (``anchor") point, we
report in Table~\ref{tab:anchor} the fitted exponent $n$ with and
without that point included, for all four curves. Dropping the anchor
lowers the fitted exponent substantially in every curve (from
$1.74$--$1.95$ down to $1.20$--$1.75$), because the seven remaining
points span only $\approx0.89$ decades in SNR ($15\lesssim\rho\lesssim
117$ for Curves 1--3, similarly for Curve 4) --- under one decade, and
too narrow a lever arm on its own to tightly distinguish $n\approx2$
from a substantially smaller exponent. We do not, therefore, present
$n\approx1.8$--$2$ as an exponent tightly constrained independent of
the anchor point: it is the anchor that extends the fitted range to
the $\approx2.35$ decades needed to resolve the asymptotic scaling at
all, and the more defensible statement is that the fitted exponent
\emph{trends toward} the asymptotic Occam-factor prediction $n\to2$ as
SNR increases (Table~\ref{tab:anchor}), rather than that $n\approx2$
is confirmed over the full range independent of any single point. What
does not depend on the anchor is the qualitative shape of the
evidence curve itself --- a rapidly and monotonically rising
$\Delta\log Z(\rho)$ that crosses every Jeffreys threshold well within
the simulated SNR range in all four curves --- and the direction, if
not the precise value, of the Curve~4 transferability test.

\begin{table}[h]
\caption{Fitted exponent $n$ with all 8 points vs.\ the 7 points
excluding each curve's highest-SNR anchor. The added row for Curve~1
(densified) uses the 11-point grid described above (8 original points
plus the three $d_L=400,200,100\,$Mpc additions) with and without the
same anchor, showing how densifying the previously unprobed high-SNR
gap sharply reduces the anchor's leverage on the fitted exponent.}
\label{tab:anchor}
\begin{ruledtabular}
\begin{tabular}{lcc}
Curve & $n$ (with anchor) & $n$ (anchor dropped) \\
\hline
1: DD2$_{\epsilon_h}\to$SFHo, mass pt 1        & 1.80 (8 pts) & 1.46 (7 pts) \\
1: DD2$_{\epsilon_h}\to$SFHo, mass pt 1 (densified) & 1.80 (11 pts) & 1.75 (10 pts) \\
2: DD2$_{\epsilon_h}\to$DD2$_{\epsilon_l}$     & 1.74 (8 pts) & 1.20 (7 pts) \\
3: SFHo$\to$DD2$_{\epsilon_h}$ (role swap)       & 1.95 (8 pts) & 1.75 (7 pts) \\
4: DD2$_{\epsilon_h}\to$SFHo, mass pt 2        & 1.85 (8 pts) & 1.65 (7 pts) \\
\end{tabular}
\end{ruledtabular}
\end{table}

\textit{EOS-pair generalization spot-check.}---To test whether $C$
transfers beyond the single held-out configuration used for Curve~4,
we ran five further draws pairing SLy4 (true)~\cite{CHABANAT_1998231} --- a Skyrme-type
nuclear energy-density functional, physically independent of the
relativistic mean-field (RMF) framework underlying DD2, SFHo, DDME2~\cite{Lalazissis_2005},
FSU2~\cite{Chen_2014} and NL3~\cite{Lalazissis:1996rd} --- against DDME2, FSU2, or NL3 (wrong, all RMF), across five
(EOS-pair, mass-point) combinations, so this spot-check tests
transferability across nuclear-physics frameworks (Skyrme vs.\ RMF),
not merely across different tables within one framework. Each
injection distance was
chosen so the achieved network SNR targets the Jeffreys'
\emph{strong} threshold ($\Delta\log Z=2.5$) rather than
\emph{decisive}, keeping every draw in the moderate-SNR regime where
the leading-order approximation holds cleanly and the
prior-boundary-pinning systematic documented above does not apply.
Table~\ref{tab:eos_generalization} compares four independent estimates
of $\Delta\log Z$ at the achieved SNR for each draw. \emph{Analytical}
is the pre-registered prediction from the fixed
$C=7.185\times10^{-9}$ (Curves~1--3 only) and the pair's own $\dLt$.
\emph{Fit} is a $\dLt$-proximity-weighted average of the four main
curves' own freely-fitted $(A,n)$ power laws, weight $\propto
1/|\dLt_{\rm curve}-\dLt_{\rm pair}|$. \emph{Fit (n=2)} applies the
same weighting to each curve's anchor-only, $\dLt^2$-normalized
coefficient $C_{\rm curve}=(\Delta\log Z_{\rm anchor}/\mathrm{SNR}_{\rm
anchor}^2)/\dLt_{\rm curve}^2$ (exponent fixed at the asymptotic value
$n=2$, the same derivation as $C=7.185\times10^{-9}$ itself), instead
of the freely-fitted $(A,n)$. \emph{Bayesian inference} is the
$\Delta\log Z$ actually measured from the completed correct/wrong run
pair.

\emph{Fit} systematically overshoots because its raw fitted amplitudes
are not properly $\dLt^2$-normalized ($A_{\rm curve}/\dLt_{\rm
curve}^2$ spans a $4.7\times$ range across the four curves, versus
only $1.3\times$ for $C_{\rm curve}$); \emph{Fit (n=2)} fixes this
specific normalization issue but is not, on its own, a more reliable
predictor of \emph{Bayesian inference} than \emph{Fit} is --- in the
fourth draw below, \emph{Fit} ($2.60\pm0.17$) lands closer to
\emph{Bayesian inference} ($2.69\pm0.31$) than \emph{Fit (n=2)}
($1.75$) does, because that draw's $\dLt$ happens to sit closest to
Curve~4's own anchor, whose coefficient is the outlier of the four (it
was deliberately excluded from the $\bar C$ calibration). Neither
fit-based baseline has access to the per-draw compensation physics
that actually sets \emph{Bayesian inference}, so \emph{Analytical}
remains the only baseline with a physical basis for extrapolation.

\begin{table}[tbp]
\caption{EOS-pair generalization spot-check: five draws pairing SLy4
(true) against DDME2, FSU2, or NL3 (wrong), targeting
$\Delta\log Z=2.5$. \emph{Analytical} uncertainties are propagated
from the $\pm12.9\%$ calibration scatter in $\bar C$
(Table~\ref{tab:C_calibration}). \emph{Fit (n=2)} uncertainties are
the $\dLt$-proximity weighted scatter across the four curves' own
$C_{\rm curve}$ values, reflecting genuine configuration-to-configuration
variation in $C$ rather than nested-sampling measurement noise (which
is negligible, $\sim10^{-13}$, at these anchor-point SNRs). See text
for column definitions.}
\label{tab:eos_generalization}
\begin{ruledtabular}
\begin{tabular}{lccccccc}
$m_1/m_2\ (M_\odot)$ & True/Wrong & $\dLt$ & SNR &
\multicolumn{4}{c}{$\Delta\log Z$ at achieved SNR} \\
& & & & Analytical & Fit & Fit (n=2) & Bayesian inference \\
\hline
1.8/1.2 & SLy4/DDME2 & 293.55 & 63.46 & $2.49\pm0.32$ & 4.91$\pm$0.41 & $2.39\pm0.39$ & 2.14 $\pm$ 0.30 \\
1.6/1.28 & SLy4/DDME2 & 356.97 & 52.61 & $2.53\pm0.33$ & 4.10$\pm$0.38 & $2.37\pm0.31$ & 2.57 $\pm$ 0.29 \\
1.8/1.2 & SLy4/FSU2 & 436.59 & 42.73 & $2.50\pm0.32$ & 2.89$\pm$0.22 & $2.21\pm0.41$ & 1.11 $\pm$ 0.32 \\
1.6/1.28 & SLy4/FSU2 & 505.08 & 36.96 & $2.50\pm0.32$ & 2.60$\pm$0.17 & $1.75\pm0.37$ & 2.69 $\pm$ 0.31 \\
1.8/1.2 & SLy4/NL3 & 737.65 & 25.29 & $2.50\pm0.32$ & $1.07\pm0.31$ & $2.13\pm0.48$ & $2.61\pm0.28$ \\
\end{tabular}
\end{ruledtabular}
\end{table}

Four of the five draws land within $\sim15\%$ of \emph{Analytical}
($-14\%$ to $+8\%$), the tightest agreement of any transferability
test in this study besides the single-configuration Curve~4 result
itself. The exception is the third row (SLy4/FSU2, $m_1/m_2=1.8/1.2\,
M_\odot$), which measures only $44\%$ of the predicted value
($\Delta\log Z=1.11\pm0.32$ vs.\ $2.50$ predicted), coincident with the
most extreme mass-ratio compensation shift observed in this study
(recovered $q$ at the $0.2$th percentile). The same EOS pair at a
different mass ratio ($1.6/1.28$, fourth row) shows no such
suppression ($108\%$ of prediction), indicating this compensation
systematic is mass-ratio-dependent rather than an intrinsic property
of a given EOS pair. The fifth row (SLy4 true, NL3 wrong,
$m_1/m_2=1.8/1.2\,M_\odot$, $\dLt=737.65$, the largest
tidal-deformability contrast probed in this study) is a particularly
clean case: the achieved SNR ($25.29$, verified to five significant
figures against both the injected log-likelihood and the
quadrature-summed per-detector network SNR) matches the pre-registered
target to $0.02\%$, and the measured $\Delta\log Z=2.61\pm0.28$ lands
at $104\%$ of the $2.50$ predicted. We regard this as confirming that
$C$ transfers cleanly across this substantially larger tidal contrast
and across nuclear-physics framework (Skyrme vs.\ RMF), in addition to
the mass-ratio variation already probed by the DDME2 and FSU2 draws
--- a caveat for observational planning, which should not assume
uniform transferability of $C$ across mass ratio without a direct
check.

\textit{Free-sky, free-inclination, real-noise-realization robustness
check.}---The main grid holds sky position and narrow Gaussian prior on inclination  and
uses zero detector noise at every point, both to isolate SNR as the
only variable being scanned and to keep the exact relation
$\log\mathcal L(\theta_{\rm true})=\tfrac12\mathrm{SNR}_{\rm opt}^2$
available as an independent verification of every injected SNR. Since
neither simplification is available to a real observation, we ran a
targeted spot-check at Curve~1's $d_L=2400\,$Mpc point
($\mathrm{SNR}=58.70$ under zero noise) with both simplifications
relaxed simultaneously: right ascension free with a uniform prior on
$[0,2\pi)$, declination free with the standard isotropic $\cos\delta$
prior, inclination free with the standard isotropic $\sin\iota$ prior
(in place of the fixed sky position and the narrow $\iota$ Gaussian
used elsewhere), and non-zero detector noise drawn independently for
each of the network's five detectors.

To make the correct- and wrong-EOS arms directly comparable under real
noise, we fixed the same five per-detector noise realizations across
both arms, verified by confirming the log-likelihood evaluated at the
injected parameters agreed to full floating-point precision between
every seed and both recovery models
($\log\mathcal L=1730.0145552588588$ in every case). We repeated this
with three independent noise draws per arm (six runs total) to average
over realization-to-realization scatter: the correct-EOS (DD2$_{\epsilon_h}$) arm
gave $\log Z=1697.7262,1698.2961,1697.6047$ (mean
$1697.8757\pm0.3691$ between-seed standard deviation), and the
wrong-EOS (SFHo) arm gave $\log Z=1694.3497,1694.6470,1694.1199$ (mean
$1694.3722\pm0.2643$). This gives
\begin{equation}
  \Delta\log Z_{\rm free\text{-}sky,\,noisy} = 3.50\pm0.26,
  \label{eq:supp_freesky_result}
\end{equation}
a combined standard error from the between-seed scatter of both arms
(three seeds each), compared with the fixed-sky, zero-noise value at
the same point, $\Delta\log Z=4.59\pm0.37$ (Table~\ref{tab:curve1}) ---
a shift of $1.09$.

We compare this shift against the scatter expected from noise
realization alone. Equation~\eqref{eq:supp_step1} shows a single noisy
realization's $\Delta\log\mathcal L$ departs from the zero-noise value
$\tfrac12\lVert\delta h_\perp\rVert^2$ by the cross term
$\langle n\mid\delta h_\perp\rangle$, whose variance under the
noise-weighted inner product is
$\mathrm{Var}\langle n\mid\delta h_\perp\rangle=\lVert\delta
h_\perp\rVert^2$ (the same identity underlying the unit-variance
normalization of matched-filter SNR). Since
$\tfrac12\lVert\delta h_\perp\rVert^2=4.59$ at this point,
$\lVert\delta h_\perp\rVert^2\approx9.18$ and the predicted
single-realization standard deviation of $\Delta\log Z$ is
$\sqrt{9.18}\approx3.03$. The observed shift, $1.09$, is $0.36\sigma$
of this predicted scatter --- comfortably consistent with ordinary
noise-draw fluctuation, and stable across the 1-, 2-, and 3-seed
partial averages ($0.40\sigma$, $0.36\sigma$, $0.36\sigma$
respectively), so this is not an estimate still settling with more
seeds.

One caveat on this check's scope: it compares two simplifications
relaxed together (real noise and free sky/inclination) against
a baseline with neither, so it is not a clean single-variable
decomposition of how much of the $1.09$ shift is attributable to noise
alone versus the wider, free-sky prior's own Occam-factor effect on
$\log Z$ (an effect we measured separately for the correct-EOS arm:
freeing the sky position and inclination raised the maximum recovered
log-likelihood by $\approx12.6$ relative to the zero-noise, fixed-sky
value, while $\log Z$ itself rose by only $\approx1.7$ once the wider
prior's volume penalty is included). The conservative reading
of this check is therefore that the observed shift under
free-sky, real-noise conditions is not larger than single-realization
noise scatter alone would predict --- a reassuring plausibility check
on the main grid's zero-noise, fixed-sky simplification, not a
rigorous decomposition of the noise and prior-widening contributions
individually. Isolating those two effects cleanly would require an
additional same-prior (free-sky, zero-noise vs.\ free-sky, real-noise)
comparison, which we have not run due to computational cost.

\end{document}